\documentclass[11pt]{revtex4-2}

\usepackage{hyperref}
\usepackage{braket}
\usepackage{amsmath,amsfonts,amssymb}
\hypersetup{dvips,dvipdfm,colorlinks=true,urlcolor=magenta,filecolor=magenta,linktoc=page,citecolor=red,linkcolor=blue,bookmarks=true}
\usepackage{graphicx,epstopdf}

\begin{document}
	\title{ \textbf{Classical and Quantum cosmology for two scalar field Brans-Dicke type theory: A Noether Symmetry approach }}
	\author{ Shriton Hembrom$^1$\footnote {shritonju@gmail.com}Roshni Bhaumik$^1$\footnote {roshnibhaumik1995@gmail.com} Sourav Dutta$^2$\footnote {sduttaju@gmail.com} Subenoy Chakraborty$^1$\footnote {schakraborty.math@gmail.com}}
	\affiliation{$^1$Department of Mathematics, Jadavpur University, Kolkata-700032, West Bengal, India\\$^2$Department of Mathematics, Dr. Meghnad Saha College, Itahar, Uttar Dinajpur-733128, West Bengal, India.}

	\begin{abstract}
		 The paper deals with a cosmological model containing two scalar fields which can be considered as an extension of the Brans-Dicke scalar field model. Due to highly coupled non linear field equations, Noether Symmetry analysis has been imposed and as a result the Lagrangian as well as the field equations become much simple in form to have the classical solutions. The relevant cosmological parameters are analyzed graphically. Finally, quantum cosmology has been studied by constructing the Wheeler-DeWitt equation and the solution of this second order partial differential equation has been done using this symmetry analysis.   
	\end{abstract}
\maketitle

\section{Introduction}

In cosmology, to describe the evolution of the Universe particularly at the early inflationary era and at the late era of accelerated expansion, the scalar field has a significant role \cite{r1, r2}. The scalar field can be considered in the gravitational action integral either with a minimal coupling between the scalar filed and the gravity or with a non--minimal coupling between them due to Mach's principle. The simplest model where gravity and scalar filed are minimally coupled \cite{r3, r4, r5} is the quintessence model while Brans-Dicke (BD) theory \cite{r6} is the common example of the other. The constant $w_{BD}$, known as the BD parameter, characterizes the theory in the sense that for $w_{BD}$ implies the significant role of the scalar field while for large $w_{BD}$ the major contribution to the dynamics comes from tensor part. Interestingly, BD theory and Einstein theory are two distinct theory in the sense that $w_{BD} \rightarrow \infty$ does not recover the usual Einstein gravity. Other examples of scalar field theory are O'Hanlon theory (a particular case of BD theory) and Galileon theory \cite{r7, r8, r9}. It is to be noted that these theories belong to a family of  scalar-tensor theory \cite{r10} termed as  Horndeski
theory \cite{r11}.\\

In the context of higher--order alternative theories of gravity, the scalar field identifies the degrees of freedoms. In \cite{r12}, a higher order gravity theory was considered where action function contains Ricci scalar and its first and second order derivatives and they have found that the theory is equivalent to a two scalar field theory of which one is the usual BD scalar field while the other scalar field is minimally coupled to gravity \cite{r13, r14, r15, r16, r17, r18}. In the present work we shall consider such two scalar field cosmology in Jordan frame where the choice of the interaction between the two scalar fields are (i) in the potential part and (ii) in the kinetic part of the action integral. It is to be noted that in an Einstein frame by considering a conformal transformation the above two models may be identified as a quintom model and chiral cosmological model (i.e., $\sigma$ model) respectively. However, for cosmological prediction, exact analytical solutions of the field equations are essential and it is very hard to find them due to highly coupled and non--linear nature of the differential equations. The theory of symmetries of differential equations and associated conserved quantities are very helpful to simplify the field equations or even solve them in some cases.

The geometric symmetries of the space time have a crucial role for investigating any physical problem. Precisely, the Noether point symmetry has an extra advantage over other geometric symmetries due to the presence of a conserved quantity termed as Noether charge. Also for distinguishing different similar physical processes this conserved charge can be considered as a selection criterion. Further, the Noether integral can be applied either to have the integrability of the system or at-least to simplify the system to a great extend. Moreover, it is possible to have self--consistency of any phenomenological physical model or constraining physical parameters involved by the Noether symmetry analysis. The plan of the paper is as follows:\\
We have discussed the basic concept of coupled BD theory in Section II, Section III gives the general idea about the Noether symmetry analysis. The application of this present work using Noether symmetry analysis has been discussed in Section IV and the solution of the model is presented in Section V. Section VI deals with the solution of wave function using Wheeler DeWitt equation. Finally, we draw our brief review of this model in Section VII.

~~~~~~~~~~~~~~~~~~~~~~~~~~~~~~~~~~~~~~~~~~~~~~~~~~~~~~~~~~~~~~~~~~~~~~~~~~~~~~~~~~~~~~~~~~~~~~~~~~~~~~~~~~~~~~~~~~~~~~~~~~~~~~~~~~~~~~~~~~~~~~~~~~~~~~~~~~~~~~~~~~~~~~~~~~~~~~~~~~~~~~~~~~~~~~~~~~~~~~~~~~~~~~~~~~~~~~~~~~~~~~~~~~~~~~~~~~~~~~~~~~~~~~~~~~~~~~~~~~~~~~~~~~~~~~~~~~~~~~~~~~~~~~~~

 \section{Basic Equations for coupled Brans-Dicke scalar field cosmology}
 The action for usual Brans-Dicke scalar field is given by \cite{r6, r10, r19} 
 \begin{equation}
 A_{BD}=\int d^4x \sqrt{-g}\bigg[\frac{1}{2}\phi R-\frac{1}{2} \frac{w_{BD}}{\phi} g^{\mu \nu}\phi_{;\mu}\phi_{;\nu}+L_m \bigg] \label{e1}
 \end{equation}
where as usual $\phi$ is the BD scalar field, $w_{BD}$ is the BD coupling parameter and $L_m$ is the matter source Lagrangian. As a further generalization let us consider another scalar field $\psi(x^k)$ which is either minimally coupled to BD scalar field or coupled to the BD field in a non--minimal way. So the above BD action integral modifies to 
  \begin{equation}
 A_{MBD}=\int d^4x \sqrt{-g}\bigg[\frac{1}{2}\phi R-\frac{1}{2} \frac{w_{BD}}{\phi} g^{\mu \nu}\phi_{;\mu}\phi_{;\nu}-\frac{\epsilon}{2}g^{\mu \nu}\psi_{;\mu}\psi_{;\nu}-V(\phi, \psi)\bigg] \label{e2}
 \end{equation}
 for the minimally coupled scalar field $\psi$ while the action takes the form 
 \begin{equation}
 A_{NMBD}=\int d^4x \sqrt{-g}\bigg[\frac{1}{2}\phi R-\frac{1}{2} \frac{w_{BD}}{\phi} g^{\mu \nu}\phi_{;\mu}\phi_{;\nu}-\frac{\epsilon}{2} \phi~ g^{\mu \nu}\psi_{;\mu}\psi_{;\nu}-V(\phi, \psi)\bigg] \label{e3}
 \end{equation}
 for non-minimal coupling between the two scalar fields. Here $\epsilon=\pm 1$ identifies the scalar field $\psi$ to be quintessence or phantom in nature.
 
 The above two action integrals in Jordan frame are quite distinct and it is intersecting to show the equivalent conformal theories in the Einstein frame. Now, under the conformal transformation $\bar{g}_{ij}=\phi g_{ij}$, the action (\ref{e1}) for the BD theory becomes
 \begin{equation}
 \bar{A}_{BD}=\int d^4x \sqrt{-g}\bigg[\frac{1}{2} R-\frac{1}{2}  g^{\mu \nu}\Phi_{;\mu}\Phi_{;\nu}+\bar{L}_m \bigg] \label{e1.1}
 \end{equation}
 Here $\Phi=\Phi(\phi(x^k))$ can be considered as a minimally coupled scalar field and $\bar{L}_m$ is the equivalent Lagrangian due to conformal transformation. Similarly, the actions (\ref{e2}) and (\ref{e3}) for the BD theory with minimally or non-minimally coupled second scalar field with the above conformal transformation take the form as
 \begin{equation}
 \bar{A}_{MBD}=\int d^4x \sqrt{-g}\bigg[\frac{1}{2} R-\frac{1}{2}  g^{\mu \nu}\Phi_{;\mu}\Phi_{;\nu}-\frac{\epsilon}{2}e^{k\Phi}g^{\mu \nu}\psi_{;\mu}\psi_{;\nu}-\bar{V}(\Phi, \psi)\bigg] \label{e2.2}
 \end{equation}
 and
 \begin{equation}
 \bar{A}_{NMBD}=\int d^4x \sqrt{-g}\bigg[\frac{1}{2} R-\frac{1}{2} g^{\mu \nu}\Phi_{;\mu}\Phi_{;\nu}-\frac{\epsilon}{2}~ g^{\mu \nu}\psi_{;\mu}\psi_{;\nu}-\bar{V}(\Phi, \psi)\bigg] \label{e3.3}
 \end{equation}
 with $\bar{V}$, the form of $V$ after the conformal transformation and $k=k(w_{BD})$. In equations (\ref{e2.2}) and (\ref{e3.3}) we have the actions for two scalar fields which are coupled in the kinetic part for equation (\ref{e2.2}) while there is no interaction among the scalar fields for the action (\ref{e3.3}). In cosmology the action (\ref{e2.2}) is known as chiral cosmology and is related to the description of the inflationary era in the form of hyperinflation, $\alpha$--attractors etc. while the action (\ref{e3.3}) corresponds to quintom model for the description of dark energy in late phase and may cross the cosmological constant boundary. One may note that even in simplest FLRW space--time model the field equations for both the above models are non--linear and coupled in nature. So to obtain exact cosmological solution Noether symmetry analysis will be very much appropriate.\\

 By varying the action (\ref{e1}) with respect to the metric tensor, one can obtain the modified Einstein field equations as
 \begin{equation}
 \phi G_{\mu \nu}= \frac{w_{BD}}{\phi^2}\bigg(\phi_{;\mu}\phi_{;\nu}-\frac{1}{2}g_{\mu \nu}g^{k \lambda}\phi_{;k}\phi_{;\lambda}\bigg)-\frac{1}{\phi}\bigg(g_{\mu \nu}g^{k \lambda}\phi_{;k \lambda}-\phi_{;\mu}\phi_{;\nu}\bigg)-g_{\mu \nu}\frac{V(\phi)}{\phi}+\frac{1}{\phi}T_{\mu \nu}
 \label{e4}
 \end{equation}
 Similarly, one can obtain the second order differential equation for the action (\ref{e2}) by varying the scalar field $\phi(x^k)$ and $\psi(x^k)$ as
 \begin{equation}
 g^{r \lambda} \phi_{;r \lambda}-\frac{1}{2\phi}g^{\mu \nu}\phi_{;\mu}\phi_{;\nu}+\frac{\phi}{2w_{BD}}(R-2V_{,\phi})=0
 \label{e5}
 \end{equation}
 and
 \begin{equation}
 g^{r \lambda} \psi_{;r \lambda}+V_{,\psi}=0
 \label{e6}
 \end{equation}
 and for action (\ref{e3}) as
 \begin{equation}
 g^{r \lambda} \phi_{;r \lambda}-\frac{1}{2\phi}g^{\mu \nu}\phi_{;\mu}\phi_{;\nu}+\frac{\epsilon \phi}{w_{BD}}g^{\mu \nu}\psi_{;\mu}\psi_{;\nu}+\frac{\phi}{2w_{BD}}(R-V_{,\phi})=0
 \label{e7}
 \end{equation}
 and
 \begin{equation}
 g^{r \lambda} \psi_{;r \lambda}+\frac{1}{\phi}g^{\mu \nu}\phi_{;\mu}\phi_{;\nu}+V_{,\psi}=0
 \label{e8}
 \end{equation}
 \section{Brief Idea About Noether Symmetry}
 According to the Mathematician Emmy Noether, there are some conserved quantity associated  to any physical system provided the Lagrangian of the system is invariant with respect to the Lie derivative along an appropriate vector filed. This is known as Noethers first theorem.\\
 
 If $L(q^{\alpha}(x^i), \dot{q}^{\alpha}(x^i))$ be a point like Lagrangian of the physical system then the Euler--Lagrange equation can be written as \cite{r20, r21, r22, r23, r24, r25, r26}
 \begin{equation}
 \partial_j\bigg(\frac{\partial L}{\partial \partial_j q^{\alpha}}\bigg)=\frac{\partial L}{\partial q^{\alpha}}
 \label{n1}
 \end{equation}
 So,  if we contract this equation (\ref{n1}) with some unknown function $\lambda^{\alpha}(q^{\beta})$ then it simplifies to
 \begin{eqnarray}
 \lambda^{\alpha}\bigg[\partial_j\bigg(\frac{\partial L}{\partial \partial_j q^{\alpha}}\bigg)-\frac{\partial L}{\partial q^{\alpha}}\bigg]&=&0\nonumber\\
\mbox{i.e.,}~~ \lambda^{\alpha}\frac{\partial L}{\partial q^{\alpha}}+(\partial_{j}\lambda^{\alpha})\left(\frac{\partial L}{\partial\partial_{j}q^{\alpha}}\right)&=&\partial_{j}\left(\lambda^{\alpha}\frac{\partial L}{\partial\partial_{j}q^{\alpha}}\right)\label{n2}
 \end{eqnarray}
 and the Lie derivative of the Lagrangian can be defined as 
 \begin{equation}
 \mathcal{L}_{\overrightarrow{X}}L=\lambda^{\alpha}\frac{\partial L}{\partial q^{\alpha}}+(\partial_{j}\lambda^{\alpha})\frac{\partial L}{\partial\left(\partial_{j}q^{\alpha}\right)}=\partial_{j}\left(\lambda^{\alpha}\frac{\partial L}{\partial\partial_{j}q^{\alpha}}\right)\label{n3}
 \end{equation}
 Here the vector $\overrightarrow{X}$ represents the infinitesimal generator of the Noether symmetry and is defined as
 \begin{equation}
 \overrightarrow{X}=\lambda^{\alpha}\frac{\partial}{\partial q^{\alpha}}+\left(\partial_{j}\lambda^{\alpha}\right)\frac{\partial}{\partial\left(\partial_{j}q^{\alpha}\right)}\label{n4}
 \end{equation}
 Now, according to Noethers theorem if the Lie derivative of the Lagrangian along the infinitesimal generator (\ref{n4}) vanishes then we can say that there exist Noether symmetry for the physical system, i.e.,
 \begin{eqnarray}
 \mathcal{L}_{\overrightarrow{X}}L&=&0\nonumber\\
 \partial_{j}\left(\lambda^{\alpha}\frac{\partial L}{\partial\partial_{j}q^{\alpha}}\right)&=&0\label{n6}
 \end{eqnarray}
 So, from (\ref{n6}), one can say that associated to this symmetry criteria, there is a constant of motion which can be written as
 \begin{equation}
 Q^i=\lambda^{\alpha}\frac{\partial L}{\partial\left(\partial_{i}q^{\alpha}\right)}\label{n7}
 \end{equation}
 This  is known as Noether current or conserved current, which satisfies the condition $\partial_iQ^i=0$. But if in the Lagrangian there is no explicit time dependence, then energy function is also a constant of motion. Then the energy function of the system is
 \begin{equation}
 E=\dot{q}^{\alpha}\frac{\partial L}{\partial\dot{q}^{\alpha}}-L\label{n8}
 \end{equation}
 This energy function is also known as Hamiltonian. In the context of quantum cosmology the Noether symmetry condition can be written as
 \begin{equation}
 \mathcal{L}_{\overrightarrow{X}_H}H=0\nonumber
 \end{equation}
 where $H$ is the Hamiltonian of the given system.  Here the vector field $\overrightarrow{X}_H$ takes the form as
 $$\overrightarrow{X}_H=\dot{q}\frac{\partial}{\partial q}+\dot{p}\frac{\partial}{\partial p}$$
and for the presence of Noether symmetry the canonically conjugate momenta corresponding to the conserved current is constant, that means 
 \begin{equation}
 \pi_{l}=\frac{\partial L}{\partial \dot{q^{l}}} =\Sigma_{l}~,~~~\mbox{a constant}\label{n9}
 \end{equation}
where $l$ goes to 1 to $r$(number of symmetries). Operator version of (\ref{n9}) can be written as
 \begin{equation}
 -i\partial_{q^l}|\psi>=\Sigma_{l}|\psi>\label{n10}
 \end{equation}
 where $|\psi>$ represents the wave function of the Universe. Now, by solving equation (\ref{n10}) one can determine the oscillatory part of the wave function. The solution of equation (\ref{n10}) takes the form
 \begin{equation}
 |\psi>=\sum_{l=1}^{r}e^{i\Sigma_{l}q^{l}}|\phi(q^\sigma)>,~~\sigma<n\label{n11}
 \end{equation}
 `$\sigma$' represents the directions along which the symmetry does not exist. Thus the oscillatory part of the wave function assures the existence of Noether symmetry.
 \section{Application of Noether Symmetry in Brans--Dicke cosmological Model}
 Cosmological principle states that in the large scale structure the Universe is assumed to be homogeneous and isotropic and is described by the metric
  \begin{equation}
  ds^2=-N^2(t) dt^2+a^2(t)(dx^2+dy^2+dz^2)\label{4.1}
  \end{equation}
 Here $N(t)$ is the lapse function and $a(t)$ is the scale factor of the Universe and the usual Ricci scalar from (\ref{4.1}) can be written as
 \begin{equation}
 R=\frac{6}{N^2}\bigg[\frac{\ddot{a}}{a}+\big(\frac{\dot{a}}{a}\big)^2-\frac{\dot{a}\dot{N}}{aN}\bigg]\label{4.2}
 \end{equation}
 where the over dots indicates the derivative with respect to the cosmic time $t$.\\
 
 Now from the action (\ref{e2}), using equation (\ref{4.2}) one can obtain the point like Lagrangian of the given system
 \begin{equation}
 \mathcal{L}=\frac{1}{2N}\bigg(-6a\phi \dot{\phi}^2-6a^2\dot{a}\dot{\phi}-\frac{w_{BD}}{\phi}a^3\dot{\phi}^2-\epsilon a^3\dot{\psi}^2\bigg)+a^3NV(\phi, \psi)\label{4.3}
 \end{equation}
 By varying the above Lagrangian with  respect to the variables $(a, \phi, \psi)$ and choosing $N=1$ one can obtain the following field equations of the Brans--Dicke cosmological model as
 \begin{equation}
 3\phi H^2+2\phi \dot{H}+2H\dot{\phi}-\frac{w_{BD}}{2}\dot{\phi}^2-\frac{\epsilon}{2}\dot{\psi}^2+\ddot{\phi}+V(\phi, \psi)=0\label{4.4}
 \end{equation}
 \begin{equation}
 3\dot{H}+6H^2+3w_{BD}H\frac{\dot{\phi}}{\phi}-\frac{\epsilon}{2}\dot{\psi}^2-\frac{w_{BD}}{2}\bigg(\frac{\dot{\phi}^2}{\phi^2}-2\frac{\ddot{\phi}}{\phi}\bigg)+V_{,\phi}(\phi, \psi)=0\label{4.5}
 \end{equation}
 \begin{equation}
 \epsilon \ddot{\psi}+3\epsilon H \dot{\psi}+V_{,\psi}(\phi, \psi)=0\label{4.6}
 \end{equation}
 Now by the variation with respect to the lapse function provides the constraint equation
 \begin{equation}
 6\phi H^2+6H\dot{\phi}+\frac{w_{BD}}{\phi}\dot{\phi}^2+\frac{\epsilon}{2}\dot{\psi}^2+2V(\phi, \psi)=0\label{4.7}
 \end{equation} 
 Here $H$ is the Hubble parameter defined by $\frac{\dot{a}}{a}$. For $N=N(a, \phi, \psi)$ the above field equations (\ref{4.4}--\ref{4.7}) describe a point particle's element takes the form as
 \begin{equation}
 ds_1^2=\frac{1}{N}\bigg(-6a\phi d\phi^2-6a^2da d\phi-\frac{w_{BD}}{\phi}a^3 d\phi^2-\epsilon \phi a^3 d \psi^2\bigg)\label{4.8}
 \end{equation}
 with effective potential $V_{eff}=a^3NV(\phi, \psi)$.\\
 
 In this section our main focus to apply Noether theorem in the given Lagrangian (\ref{4.3}). So for the present three dimensional point like Lagrangian the infinitesimal generator takes the form as
 \begin{equation}
 \overrightarrow{X}=\alpha\frac{\partial}{\partial a}+\beta\frac{\partial}{\partial \phi}+\gamma\frac{\partial}{\partial \psi}+\delta\frac{\partial}{\partial N}+\dot{\alpha}\frac{\partial}{\partial \dot{a}}+\dot{\beta}\frac{\partial}{\partial \dot{\phi}}+\dot{\gamma}\frac{\partial}{\partial \dot{\psi}}\label{4.9}
 \end{equation}
 Now by using the condition (\ref{n6}) we can see that the co-efficients of the infinitesimal generator $(\alpha, \beta, \gamma, \delta)$ have to satisfy the following set of partial differential equations
 \begin{equation}
 -\frac{3\alpha \phi}{N}-\frac{3a \beta}{N}+\frac{3a \phi}{N^2}\delta-\frac{6a\phi}{N}\frac{\partial \alpha}{\partial a}-\frac{3a^2}{N}\frac{\partial \beta}{\partial a}=0\label{4.10}
 \end{equation}
 \begin{equation}
 \frac{w_{BD}a^3}{2\phi^2N}\beta+\frac{w_{BD}}{2N^2\phi}a^3 \delta-\frac{3a^2}{N}\frac{\partial \alpha}{\partial \phi}-\frac{w_{BD}}{N\phi}a^3\frac{\partial \beta}{\partial \phi}-\frac{3w_{BD}}{2\phi N}a^2\alpha=0\label{4.11}
 \end{equation}
 \begin{equation}
 -\frac{3\epsilon}{2N}a^2\alpha+\frac{\epsilon}{2N^2}a^3 \delta-\frac{\epsilon}{N}a^3\frac{\partial \gamma}{\partial \psi}=0\label{4.12}
 \end{equation}
 \begin{equation}
 \frac{-6}{N}a\alpha+\frac{3}{N^2}a^2\delta-\frac{3a^2}{N}\frac{\partial \alpha}{\partial a}-\frac{6a\phi}{N}\frac{\partial \alpha}{\partial \phi}-\frac{3a^2}{N}\frac{\partial \beta}{\partial \phi}-\frac{w_{BD}}{N\phi}a^3\frac{\partial \beta}{\partial a}=0\label{4.13}
 \end{equation}
 \begin{equation}
 -\frac{3a^2}{N}\frac{\partial \alpha}{\partial \psi}-\frac{w_{BD}}{N\phi}a^3\frac{\partial \beta}{\partial \psi}-\frac{\epsilon}{N}a^3\frac{\partial V}{\partial \phi}=0\label{4.14}
 \end{equation}
 \begin{equation}
 3\alpha a^2 N V(\phi, \psi)+a^3 \beta N\frac{\partial V(\phi, \psi)}{\partial \phi}+\gamma a^3 N \frac{\partial V(\phi, \psi)}{\partial \psi}+\delta a^3 V(\phi, \psi)=0\label{4.15}
 \end{equation}
 Hence by solving the over determined set of equations (\ref{4.10}-\ref{4.15}) by the method of separation of variable we can find that the symmetry vector (\ref{4.9}) admits the following set of Noether symmetries\\
 
 {\textbf{Case:I}}~~$X_1=\alpha_0a \frac{\partial}{\partial a}+\beta_0 \phi \frac{\partial}{\partial \phi}+\gamma_0 \psi \frac{\partial}{\partial \psi}+\delta_0N\frac{\partial}{\partial N}$, with the unknown potential $V(\phi, \psi)=v_0\phi^{\frac{k}{\beta_0}}\psi^{\frac{k_1}{\gamma_0}}$ where $\alpha_0, \beta_0, \gamma_0, \delta_0, k, k_1$ are the arbitrary constants related by the relations $3\alpha_0+\beta_0-\delta_0=3\alpha_0+2\gamma_0-\delta_0=0,~k_1=-3\alpha_0-\delta_0-k$.\\
 
 {\textbf{Case:II}}~~$X_2=\alpha'_0a \frac{\partial}{\partial a}+\beta'_0 \phi \frac{\partial}{\partial \phi}+\gamma'_0\psi \frac{\partial}{\partial \psi}$, here we have the potential function as $V(\phi, \psi)=\psi^2$, with $\alpha'_0, \beta'_0, \gamma'_0$ are the arbitrary constants related to $3\alpha'_0+\beta'_0=3\alpha'_0+2\gamma'_0=0$.\\
 
 {\textbf{Case:III}}~~$X_2=\alpha''_0a \frac{\partial}{\partial a}-\frac{6\alpha''_0}{\lambda} \frac{\partial}{\partial \psi}+3\alpha''_0N \frac{\partial}{\partial N}$, and in this case the potential function  $V(\phi, \psi)=e^{\lambda \psi}$, with $\alpha''_0, \lambda$ as the arbitrary constants.\\
 
 {\textbf{Case:IV}}~~$X_4=\alpha'''_0a \frac{\partial}{\partial a}+\beta'''_0 \phi \frac{\partial}{\partial \phi}+\gamma'''_0 \psi \frac{\partial}{\partial \psi}+\delta'_0N\frac{\partial}{\partial N}$, with the unknown potential $V(\phi, \psi)=\phi^2$ where $\alpha'''_0, \beta'''_0, \gamma'''_0, \delta'_0$ are the arbitrary constants related by the relations $2\alpha'''_0=-\beta'''_0=2\delta'_0=-2\gamma'''_0$.\\
 \section{Solution of the model}
 In this section we are trying to find the exact solution of the present model. For this purpose we choose a suitable co-ordinate transformation $(a, \phi, \psi, N)\rightarrow (u, v, w, w_1)$ in such a way that the transformed Lagrangian i.e., the Lagrangian in terms of new variables contains at least one cyclic variable.
 So, now for each case we have restricted this transformation as
 \begin{equation}
 i_{_{\overrightarrow{X}}} du=1,~i_{_{\overrightarrow{X}}} dv=0,~i_{_{\overrightarrow{X}}}dw=0,~i_{_{\overrightarrow{X}}} dw_1=0\label{5.1}
 \end{equation}
 here the transformed symmetry vector is along the direction of $u$ and perpendicular to the directions of $v, w, w_1$. $i_{_{\overrightarrow{X}}}$  in equation (\ref{5.1}) stands for the inner-product. Hence by solving equation (\ref{5.1}) one can gets the transformed Lagrangian for the following cases:\\
 
{\textbf{Case:I:}}~The relation between the old and new variables can be expressed as
 \begin{equation}
 \left.\begin{array}{lll}
 a &=& e^{-u} \\
 a^2 \phi&=&e^v\\
 a\psi&=&e^w\\
 \frac{a^3 \phi}{N}&=&e^{w_1}
 \end{array}\right\} \label{5.2}
 \end{equation}
 Hence by using (\ref{5.2}) the transformed Lagrangian takes the form
 \begin{equation}
 L_T=e^{w_1}\bigg[(3-2w_{BD})\dot{u}^2+(3-2w_{BD})\dot{u}\dot{v}-\frac{w_{BD}}{2}\dot{v}^2-\frac{\epsilon}{2}e^{2w-v}(\dot{w}^2+2\dot{w}\dot{u}+\dot{u}^2)+v_0e^{-2w_1}e^{(1+\frac{k}{2})v}e^{(4-k)w}\bigg]\label{5.3}
 \end{equation}
 where the new variable $u$ acts as a cyclic coordinate. So corresponding to this transformed Lagrangian the field equations are much simpler than the previous one. Hence the Euler Lagrange equations can be written as
 \begin{equation}
 2\dot{u}(3-2w_{BD})+\dot{v}(3-2w_{BD})-\epsilon e^{2w-v}(\dot{w}+\dot{u})=A~(constant).\label{5.4}
 \end{equation}
 \begin{equation}
 (3-2w_{BD})\ddot{u}-w_{BD}\ddot{v}+\frac{\epsilon}{2}(\dot{w}^2+2\dot{w}\dot{u}+\dot{u}^2)e^{2w-v}-v_0e^{-2w_1}(1+\frac{k}{2})e^{(1+\frac{k}{2})v}e^{(4-k)w}=0\label{5.5}
 \end{equation}
 \begin{equation}
 \frac{d}{dt}\big\{e^{2w-v}(\dot{w}+\dot{u})\big\}-\epsilon(\dot{w}+\dot{u})^2e^{2w-v}-v_0e^{-2w_1}e^{(1+\frac{k}{2})v}(4-k)e^{(4-k)w}=0\label{5.6}
 \end{equation}
 \begin{equation}
 (3-2w_{BD})\dot{u}^2+(3-2w_{BD})\dot{u}\dot{v}-\frac{w_{BD}}{2}\dot{v}^2-\frac{\epsilon}{2}(\dot{w}^2+2\dot{w}\dot{u}+\dot{u}^2)e^{2w-v}-v_0e^{-2w_1}e^{(1+\frac{k}{2})v}e^{(4-k)w}=0\label{5.7}
 \end{equation}
 Solving the above set of equations(\ref{5.4}--\ref{5.7}), one can obtain the explicit solution for the following cases:\\
 {\textbf{Subcase I:}~$w(t)=w_0$, a constant
  \begin{equation}
 \left.\begin{array}{lll}
 u(t) &=& 2c_1p\tan \big(\frac{t+c_2}{2c_1}\big)+c_3 \\
 v(t)&=&\ln\big\{\frac{\sec^2\big(\frac{t+c_2}{2c_1}\big)}{2Bc_1^2}\big\}\\
 \mbox{and}~~w_{BD}&=&\frac{3}{2}
 \end{array}\right\} \label{5.8}
 \end{equation}
 Where $p=\frac{-A}{2\epsilon Bc_1^2}e^{-2w_0},~B=\frac{2 A^2e^{-2w_0}}{\epsilon (4-k)}, A, B, c_1, c_2, c_3$ are the arbitrary constants.\\

 {\textbf{Subcase:II}}~$w(t)=lv(t), c>0$, ($l, c$ are constants)
 \begin{equation}
 \left.\begin{array}{lll}
 u(t) &=& \frac{-2l}{2l-1}\ln \bigg[\sqrt{\frac{2m}{(1-2l)c}} \sinh \bigg(\frac{(2l-1)\sqrt{c}}{2}(t+c_4)\bigg)\bigg]-\frac{A\sqrt{c}}{m\epsilon} \coth\bigg(\frac{(2l-1)\sqrt{c}}{2}(t+c_4)\bigg)+c_5 \\
 v(t)&=&\frac{2}{2l-1}\ln \bigg[\sqrt{\frac{2m}{(1-2l)c}} \sinh \bigg(\frac{(2l-1)\sqrt{c}}{2}(t+c_4)\bigg)\bigg]\\
 w(t)&=&\frac{2l}{2l-1}\ln \bigg[\sqrt{\frac{2m}{(1-2l)c}} \sinh \bigg(\frac{(2l-1)\sqrt{c}}{2}(t+c_4)\bigg)\bigg]\\
 \mbox{and}~~w_{BD}&=&\frac{3}{2}
 \end{array}\right\} \label{5.8.1}
 \end{equation}
 here  $c_4, c_5, m$ are  arbitrary constants.\\
 
 {\textbf{Subcase:III}}~$w(t)=lv(t), c<0$
 \begin{equation}
 \left.\begin{array}{lll}
 u(t) &=& \frac{-2l}{2l-1}\ln \bigg[\sqrt{\frac{2m}{(1-2l)c}} \sin \bigg(\frac{(2l-1)\sqrt{c}}{2}(t+c_6)\bigg)\bigg]-\frac{A\sqrt{c}}{m\epsilon} \cot\bigg(\frac{(2l-1)\sqrt{c}}{2}(t+c_4)\bigg)+c_7 \\
 v(t)&=&\frac{2}{2l-1}\ln \bigg[\sqrt{\frac{2m}{(1-2l)c}} \sin \bigg(\frac{(2l-1)\sqrt{c}}{2}(t+c_6)\bigg)\bigg]\\
 w(t)&=&\frac{2l}{2l-1}\ln \bigg[\sqrt{\frac{2m}{(1-2l)c}} \sin \bigg(\frac{(2l-1)\sqrt{c}}{2}(t+c_6)\bigg)\bigg]\\
 \mbox{and}~~w_{BD}&=&\frac{3}{2}
 \end{array}\right\} \label{5.8.2}
 \end{equation}

  Hence the classical cosmological solutions in the old variables for the above three cases:\\

  {\textbf{Subcase:I}}
 \begin{equation}
 \left.\begin{array}{lll}
 a(t) &=& e^{-2c_1p\tan \big(\frac{t+c_2}{2c_1}\big)-c_3} \\
 \phi(t)&=&\frac{1}{2Bc_1^2}\bigg(\sec^2\big(\frac{t+c_2}{2c_1}\big)\bigg)e^{4c_1p\tan \big(\frac{t+c_2}{2c_1}\big)+2c_3}\\
 \psi(t)&=&e^{w_0}e^{2c_1p\tan \big(\frac{t+c_2}{2c_1}\big)+c_3}
 \end{array}\right\} \label{5.9}
 \end{equation}
 
 {\textbf{Subcase:II}}
 \begin{equation}
 \left.\begin{array}{lll}
 a(t) &=& \bigg(\sqrt{\frac{2m}{(1-2l)c}} \sinh \bigg(\frac{(2l-1)\sqrt{c}}{2}(t+c_4)\bigg)\bigg)^{\frac{-2l}{1-2l}}~e^{\frac{A\sqrt{c}}{m\epsilon} \coth\bigg(\frac{(2l-1)\sqrt{c}}{2}(t+c_4)\bigg)}+c_5 \\
 \phi(t)&=&\bigg(\sqrt{\frac{2m}{(1-2l)c}} \sinh \bigg(\frac{(2l-1)\sqrt{c}}{2}(t+c_4)\bigg)\bigg)^{\frac{4l}{2l-1}}~e^{-\frac{2A\sqrt{c}}{m\epsilon} \coth\bigg(\frac{(2l-1)\sqrt{c}}{2}(t+c_4)\bigg)-2c_5}\\
 &&\bigg(\sqrt{\frac{2m}{(1-2l)c}} \sinh \bigg(\frac{(2l-1)\sqrt{c}}{2}(t+c_4)\bigg)\bigg)^{\frac{2}{2l-1}} \\
 \psi(t)&=&\bigg(\sqrt{\frac{2m}{(1-2l)c}} \sinh \bigg(\frac{(2l-1)\sqrt{c}}{2}(t+c_4)\bigg)\bigg)^{\frac{4l}{1-2l}}~e^{-\frac{A\sqrt{c}}{s\epsilon} \coth\bigg(\frac{(2l-1)\sqrt{c}}{2}(t+c_4)\bigg)+c_5}\\ &&\bigg(\sqrt{\frac{2m}{(1-2l)c}} \sinh \bigg(\frac{(2l-1)\sqrt{c}}{2}(t+c_4)\bigg)\bigg)^{\frac{2l}{2l-1}}
 \end{array}\right\} \label{5.9.1}
 \end{equation}
 
 {\textbf{Subcase:III}}
 \begin{equation}
 \left.\begin{array}{lll}
 a(t) &=& \bigg(\sqrt{\frac{2m}{(1-2l)c}} \sin \bigg(\frac{(2l-1)\sqrt{c}}{2}(t+c_6)\bigg)\bigg)^{\frac{-2l}{1-2l}}~e^{\frac{A\sqrt{c}}{m\epsilon} \cot\bigg(\frac{(2l-1)\sqrt{c}}{2}(t+c_6)\bigg)+c_7} \\
 \phi(t)&=&\bigg(\sqrt{\frac{2m}{(1-2l)c}} \sin \bigg(\frac{(2l-1)\sqrt{c}}{2}(t+c_6)\bigg)\bigg)^{\frac{4l}{2l-1}}~e^{-\frac{2A\sqrt{c}}{m\epsilon} \cot\bigg(\frac{(2l-1)\sqrt{c}}{2}(t+c_6)\bigg)-2c_7}\\
 &&\bigg(\sqrt{\frac{2m}{(1-2l)c}} \sin \bigg(\frac{(2l-1)\sqrt{c}}{2}(t+c_4)\bigg)\bigg)^{\frac{2}{2l-1}} \\
 \psi(t)&=&\bigg(\sqrt{\frac{2m}{(1-2l)c}} \sin \bigg(\frac{(2l-1)\sqrt{c}}{2}(t+c_6)\bigg)\bigg)^{\frac{4l}{1-2l}}~e^{-\frac{A\sqrt{c}}{s\epsilon} \cot\bigg(\frac{(2l-1)\sqrt{c}}{2}(t+c_6)\bigg)+c_7}\\
 &&\bigg(\sqrt{\frac{2m}{(1-2l)c}} \sin \bigg(\frac{(2l-1)\sqrt{c}}{2}(t+c_4)\bigg)\bigg)^{\frac{2l}{2l-1}}
 \end{array}\right\} \label{5.9.2}
 \end{equation}
 
{\textbf{Case: II:}} For the symmetry vector $X_2$ one can get the interrelation between the old and the new variables using equation(\ref{5.1}) as
 $$a=e^{-2u},~a^3\phi=e^v,~a^3 \psi^2=e^w$$
As a consequence, the transformed Lagrangian in new variables takes the form
$$L_T=e^v\bigg[-12\dot{u}^2+6\dot{u}(\dot{v}+6\dot{u})-\frac{w_{BD}}{2}(\dot{v}+6\dot{u})^2\bigg]-\frac{\epsilon}{8}e^w(\dot{w}+2\dot{u})^2+e^w$$
Though the above transformed Lagrangian as well as the transformed field equations are much simpler than the old variable, but still it is not possible to have an analytic solution in this case.\\

{\textbf{Case: III:}} Due to symmetry vector $X_3$, it is possible to have a transformation of the variables in the augmented space so that the system gets simplified using equation(\ref{5.1}). The interrelation between the old and the new variables are given by
$$u=\ln a,~v=\phi,~w=6\ln a+\lambda \psi,~D=\ln(\frac{a^3}{N})$$
So the transformed Lagrangian has the expression
$$L_T=e^D\bigg[-3v\dot{v}^2-3\dot{u}\dot{v}-\frac{w_{BD} \dot{v}^2}{2v}\bigg]-\frac{\epsilon}{2\lambda^2}e^D(\dot{w}-6\dot{u})^2+e^{w-D}$$
similar to case II, no analytic solution of the field equations corresponding to the above Lagrangian is  possible.\\

{\textbf{Case: IV:}} For the symmetry vector $X_4$ we have proceed as for the earlier three cases using equation(\ref{5.1}) and the interrelation between the old variable and the new variable are given by
$$a=e^{-u},~\phi=e^{2u+v}~,\psi=e^{u+w}~,N=e^{v-u-D}$$
The corresponding form of the Lagrangian in the new variable is given by 
$$L_T=e^D\bigg[-3\dot{u}^2+3\dot{u}(\dot{v}+2\dot{u})-\frac{w_{BD}}{2}(\dot{v}+2\dot{u})^2-\frac{\epsilon}{2}e^{2w-v}(\dot{u}+\dot{w})^2\bigg]+e^{3v-D}$$
Thus the solutions of the corresponding Euler-Lagrange equations for different choices for $w$ are expressed as\\

{\textbf{Subcase I:}}~$w(t)=m$, a constant
\begin{equation}
\left.\begin{array}{lll}
u(t)&=&\frac{-Be^{-2m}}{\epsilon \sqrt{M}}\ln\bigg|\sec\bigg(\frac{t+c_8}{c_9}\bigg)+\tan\bigg(\frac{t+c_8}{c_9}\bigg)\bigg|+c_{10} \\
v(t)&=&\frac{1}{2}\ln\bigg(\frac{\sec^2\big(\frac{t+c_8}{c_9}\big)}{Mc_9^2}\bigg)\\
\mbox{and}~~w_{BD}&=&\frac{3}{2}
\end{array}\right\} \label{5.9.3}
\end{equation} 

{\textbf{Subcase II:}}~$w(t)=-\frac{1}{2}v(t)$
\begin{equation}
\left.\begin{array}{lll}
u(t)&=&\ln\bigg(\frac{\sec^2\big(\frac{t+c_8}{c_9}\big)}{Mc_9^2}\bigg)^{\frac{1}{4}}-\frac{B}{Mc_9}\tan\bigg(\frac{t+c_8}{c_9}\bigg)
+c_{10} \\
v(t)&=&\ln\bigg(\frac{\sec^2\big(\frac{t+c_8}{c_9}\big)}{Mc_9^2}\bigg)^{\frac{1}{2}}\\
w(t)&=&\ln\bigg(\frac{\sec^2\big(\frac{t+c_8}{c_9}\big)}{Mc_9^2}\bigg)^{\frac{-1}{4}}\\
\mbox{and}~~w_{BD}&=&\frac{3}{2}
\end{array}\right\} \label{5.9.4}
\end{equation}

{\textbf{Subcase III:}}~$w(t)=\frac{1}{2}v(t)$
\begin{equation}
\left.\begin{array}{lll}
u(t)&=&\ln\bigg(\frac{\sec^2\big(\frac{t+c_8}{c_9}\big)}{Mc_9^2}\bigg)^{-\frac{1}{4}}-Bt+c_{10} \\
v(t)&=&\ln\bigg(\frac{\sec^2\big(\frac{t+c_8}{c_9}\big)}{Mc_9^2}\bigg)^{\frac{1}{2}}\\
w(t)&=&\ln\bigg(\frac{\sec^2\big(\frac{t+c_8}{c_9}\big)}{Mc_9^2}\bigg)^{\frac{1}{4}}\\
\mbox{and}~~w_{BD}&=&\frac{3}{2}
\end{array}\right\} \label{5.9.6}
\end{equation}
with $B, m, M=2e^{-2D}, c_8, c_9, c_{10}$ are arbitrary constants in the above solutions. Thus the classical solutions in old variables takes the form\\

{\textbf{Subcase I:}}
 \begin{equation}
\left.\begin{array}{lll}
a(t)&=&\bigg(\sec\bigg(\frac{t+c_8}{c_9}\bigg)+\tan\bigg(\frac{t+c_8}{c_9}\bigg)\bigg)^{\frac{Be^{-2m}}{\epsilon \sqrt{M}}}e^{c_{10}}\\
\phi(t)&=&\frac{1}{\sqrt{M}c_9}\sec\bigg(\frac{t+c_8}{c_9}\bigg)\bigg(\sec\bigg(\frac{t+c_8}{c_9}\bigg)+\tan\bigg(\frac{t+c_8}{c_9}\bigg)\bigg)^{\frac{-2Be^{-2m}}{\epsilon \sqrt{M}}}e^{-2c_{10}}\\
\psi(t)&=&e^m\bigg(\sec\bigg(\frac{t+c_8}{c_9}\bigg)+\tan\bigg(\frac{t+c_8}{c_9}\bigg)\bigg)^{\frac{-Be^{-2m}}{\epsilon \sqrt{M}}}e^{-c_{10}}
\end{array}\right\} \label{5.9.7}
\end{equation}

{\textbf{Subcase II:}}
\begin{equation}
\left.\begin{array}{lll}
a(t)&=&\Big(\frac{\sec^2\big(\frac{t+c_8}{c_9}\big)}{Mc_9^2}\Big)^{\frac{-1}{4}}e^{\frac{B}{Mc_9}\tan\bigg(\frac{t+c_8}{c_9}\bigg)-c_{10}}
\\
\phi(t)&=&\frac{\sec^2\big(\frac{t+c_8}{c_9}\big)}{Mc_9^2}e^{\frac{-2B}{Mc_9}\tan\bigg(\frac{t+c_8}{c_9}\bigg)+2c_{10}}\\
\psi(t)&=&e^{\frac{-B}{Mc_9}\tan\bigg(\frac{t+c_8}{c_9}\bigg)+c_{10}}
\end{array}\right\} \label{5.9.8}
\end{equation}

{\textbf{Subcase III:}}
\begin{equation}
\left.\begin{array}{lll}
a(t)&=&\bigg(\frac{\sec^2\big(\frac{t+c_8}{c_9}\big)}{Mc_9^2}\bigg)^{\frac{1}{4}}e^{Bt-c_{10} }
\\
\phi(t)&=&e^{-2Bt+2c_{10} }\\
\psi(t)&=&e^{-Bt+c_{10} }
\end{array}\right\} \label{5.9.10}
\end{equation}
To examine whether the above six set of solutions for the two different Noether symmetry vectors (and different choices for the parameters) are compatible to classical cosmology, the three leading cosmological parameters namely the scale factor, Hubble parameter and the acceleration parameter have been plotted in Fig (\ref{f1}) for case I, considering various choices for the parameters involved. Note that we have drawn the figures only for subcase I as the figures for the other two subcases are almost similar. For Case IV, the graphical representation for the scale factor and the Hubble parameter are of same nature in all the subcases and also they are similar to those in Fig (\ref{f1}). So we have not present them again. However, the variation of the acceleration parameter has some distinct character in the three subcases  and are presented in Fig (\ref{f2}). From the graph it is clear that the present cosmological model is an expanding model of the Universe with the rate of expansion gradually decreases. For case I the Universe evolves from an accelerated phase to decelerated phase then again in the accelerated era as noted in observation data. For Case IV, the model describes the evolution from decelerated era to late time accelerated epoch for subcase I and III while the subcase II corresponds to the entire evolution of the Universe starting from the earlier accelerated era of evolution.
	
 \begin{figure}[h]
 	\begin{minipage}{0.47\textwidth}
 		\centering \includegraphics[height=5cm,width=8cm]{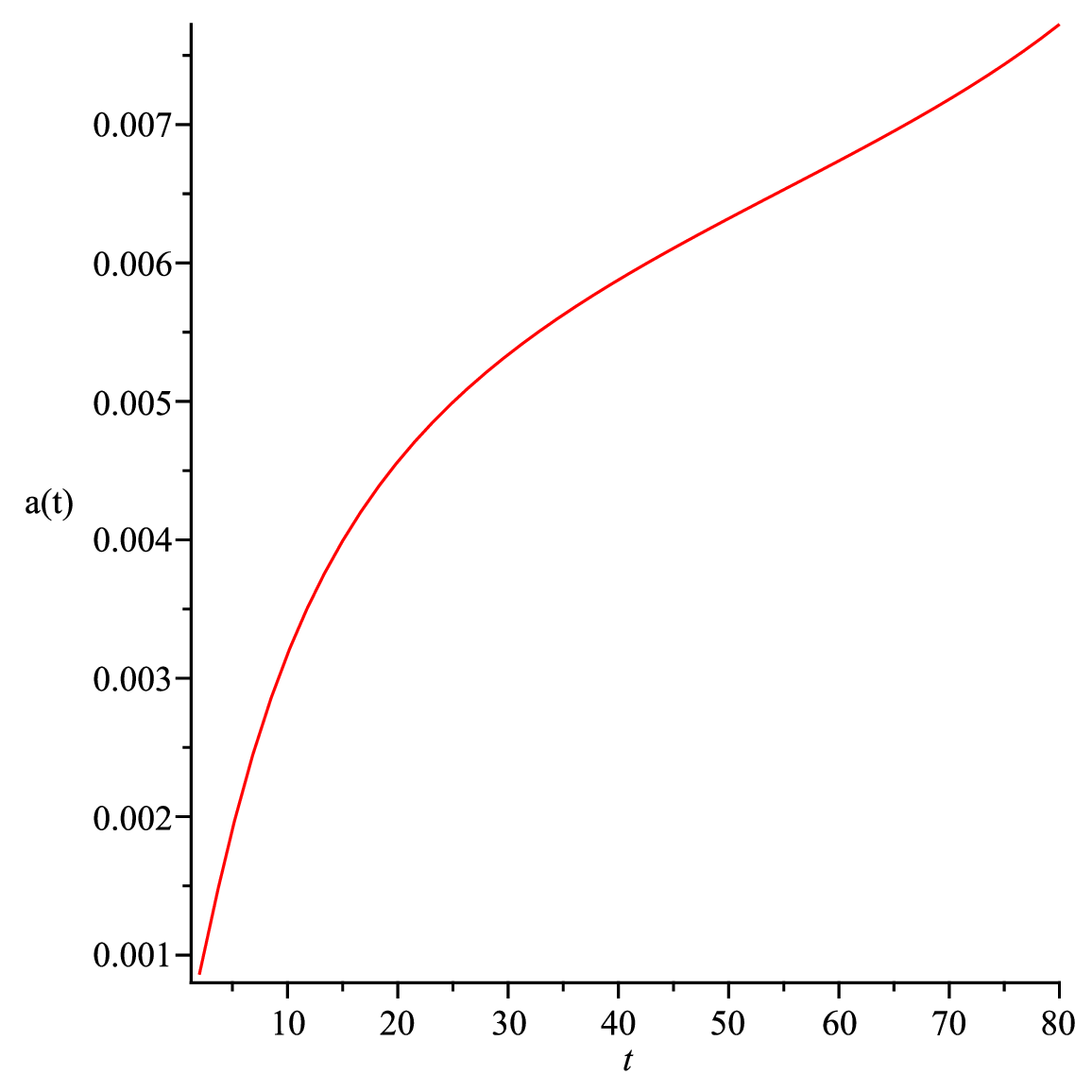}
 	\end{minipage}\hfill
 	\begin{minipage}{0.47\textwidth}
 		\centering \includegraphics[height=5cm,width=8cm]{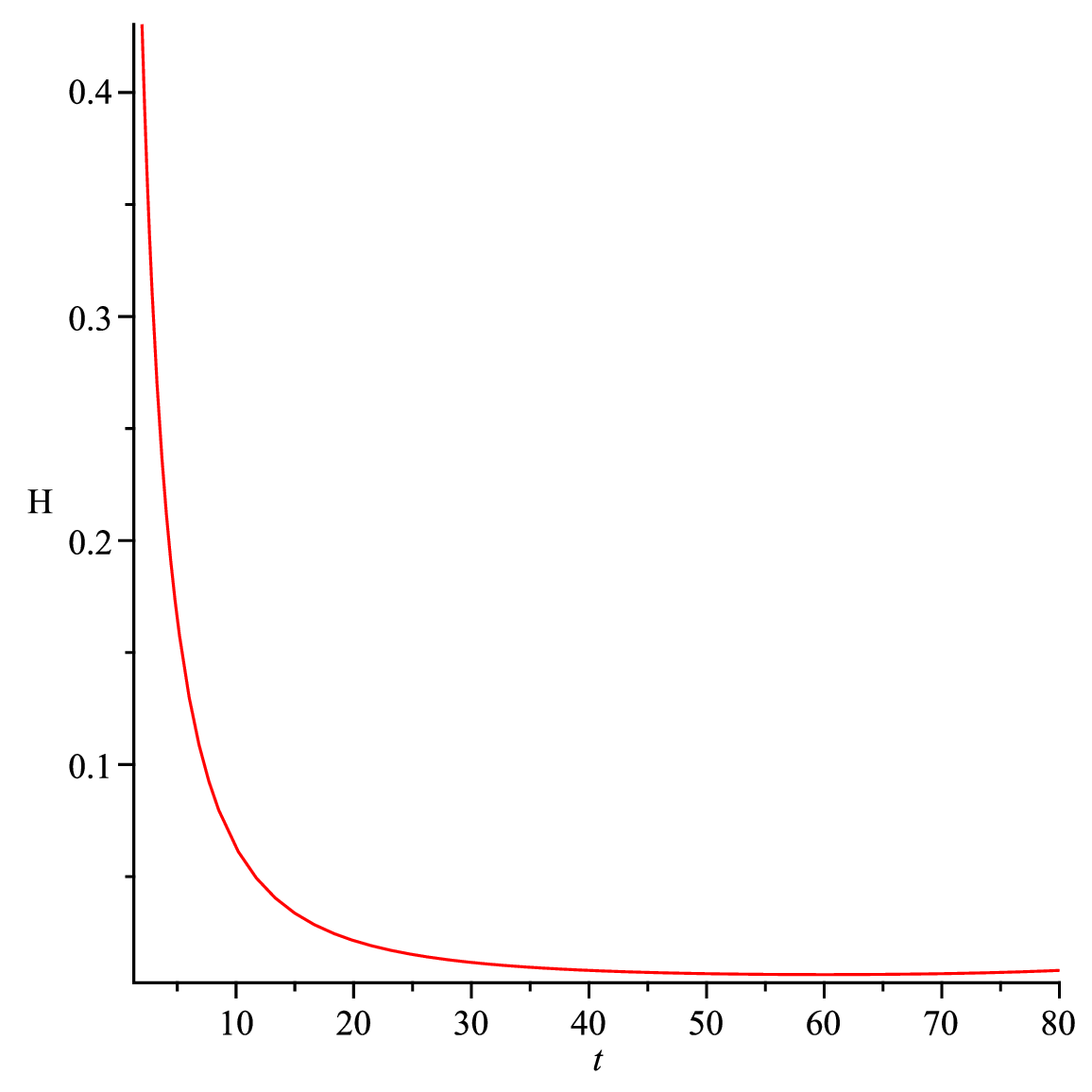}
 	\end{minipage}
 	\begin{minipage}{0.47\textwidth}
 		\centering \includegraphics[height=5cm,width=8cm]{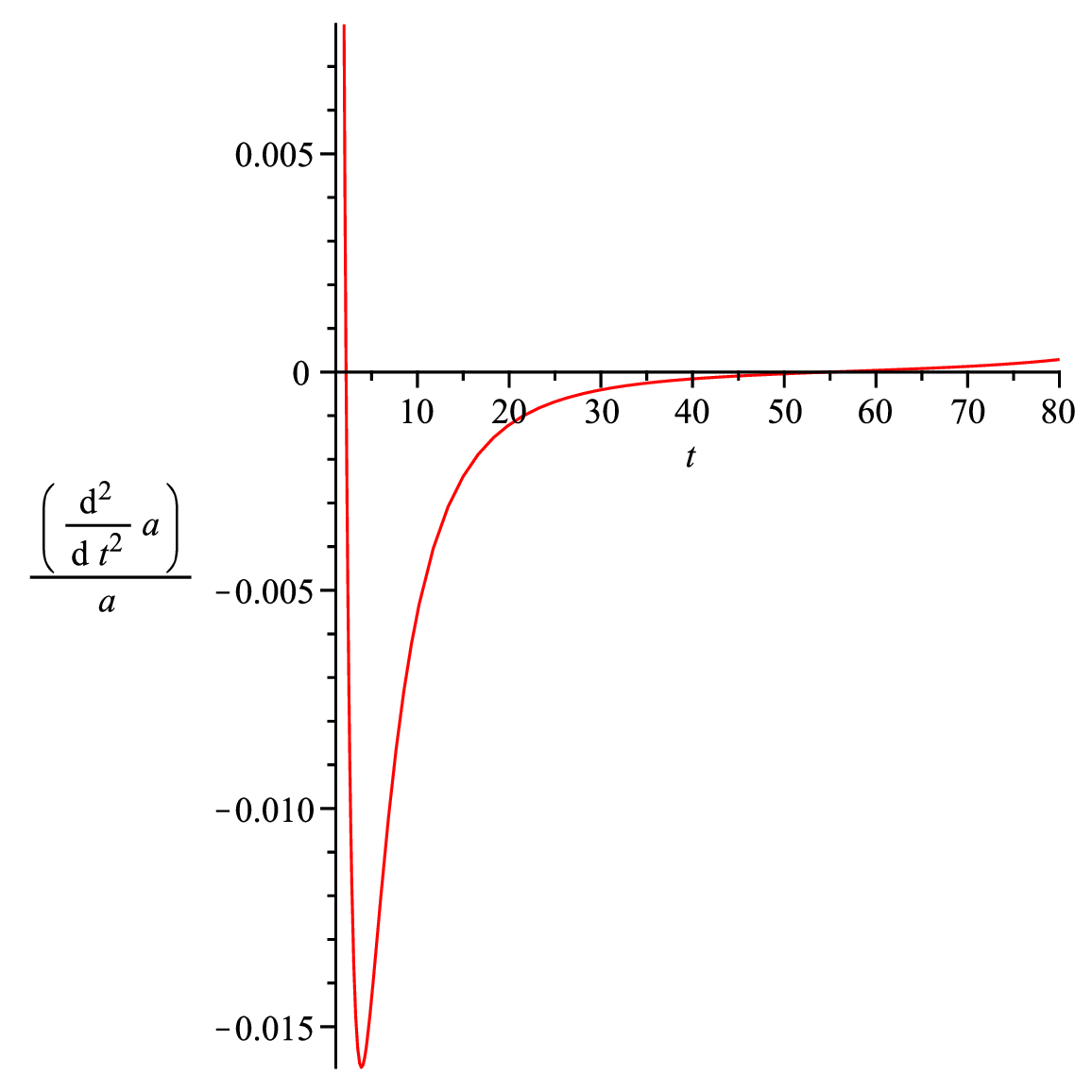}
 	\end{minipage}
 	\caption{shows the graphical representation of scale factor $a(t)$ (top left), Hubble parameter $H(t)$ (top right) and  $\frac{\ddot{a}}{a}$ (bottom) with respect to cosmic time $t$ for Case I.}\label{f1}
 \end{figure}

 \begin{figure}[h]
	\begin{minipage}{0.47\textwidth}
		\centering \includegraphics[height=5cm,width=8cm]{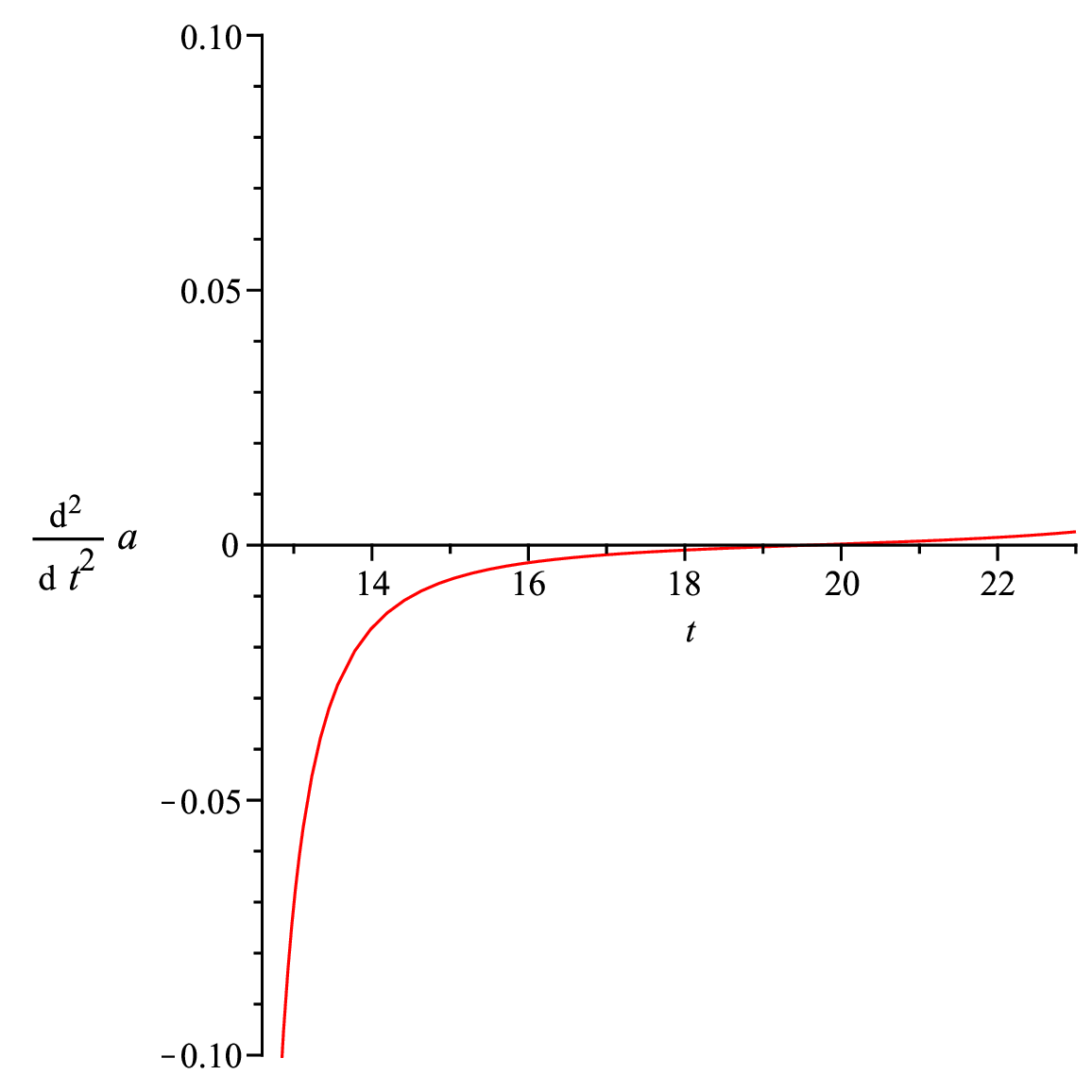}
	\end{minipage}\hfill
	\begin{minipage}{0.47\textwidth}
		\centering \includegraphics[height=5cm,width=8cm]{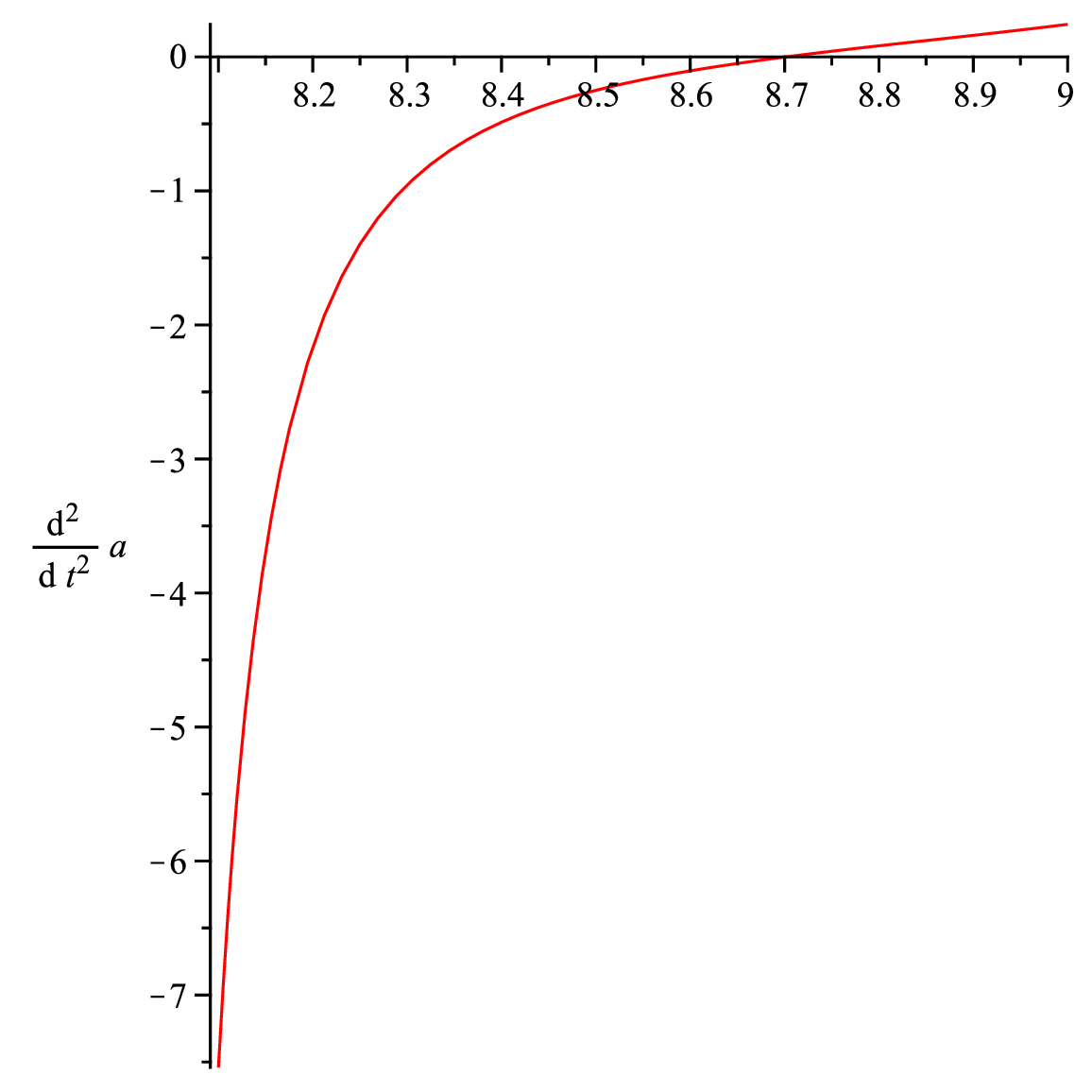}
	\end{minipage}
	\begin{minipage}{0.47\textwidth}
		\centering \includegraphics[height=5cm,width=8cm]{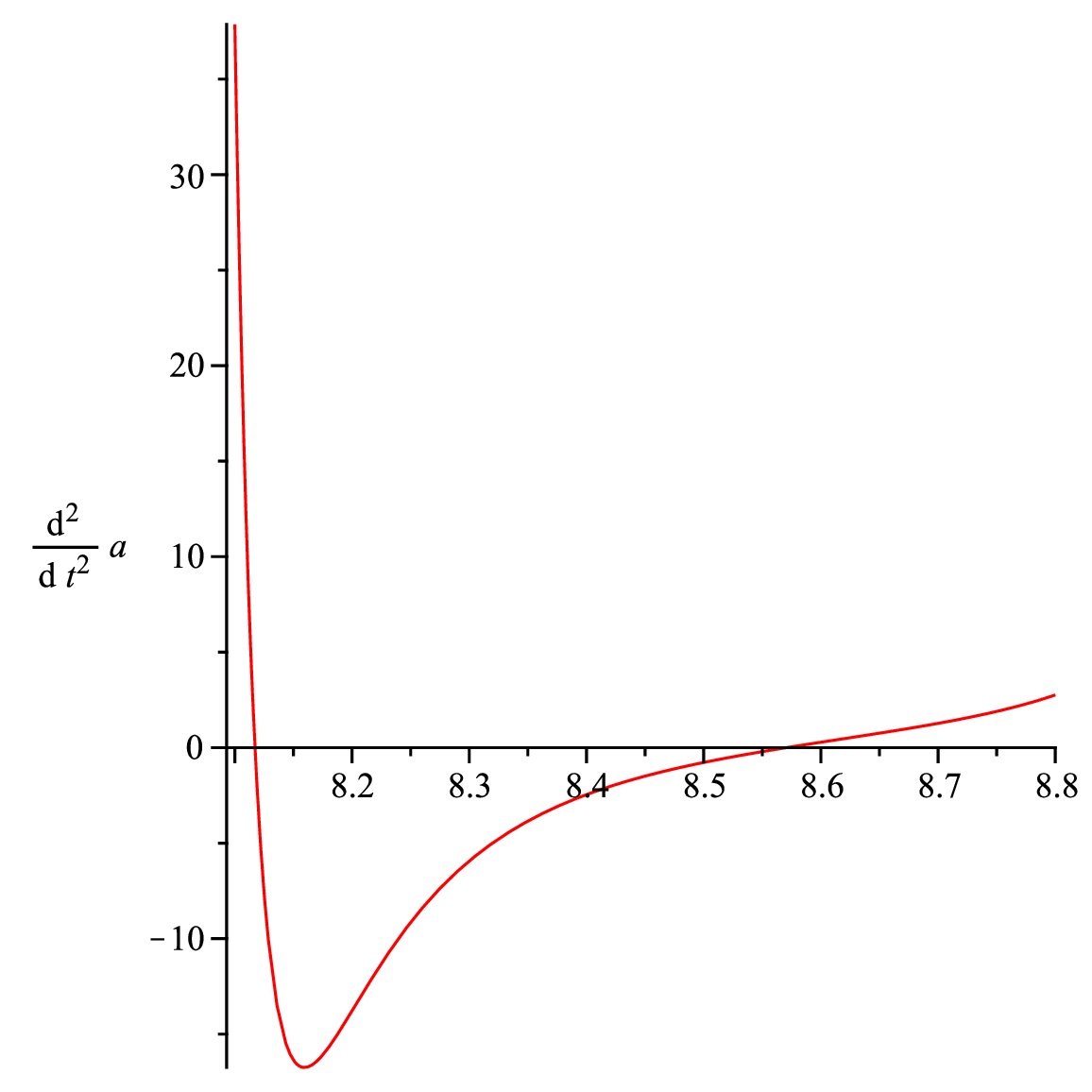}
	\end{minipage}
	\caption{Represents $\frac{\ddot{a}}{a}$ (top left) for $w=0$, $\frac{\ddot{a}}{a}$ (top right) when $w(t)=\frac{1}{2}v(t)$ and  $\frac{\ddot{a}}{a}$ (bottom) for $w(t)=-\frac{1}{2}v(t)$ with respect to cosmic time $t$ for Case IV.}\label{f2}
\end{figure}
 
  \section{Solution of the wave function using Wheeler-DeWitt equation}
  
  In the above transformed Lagrangian (\ref{5.3}) the variable $u$ acts as a cyclic coordinate, so in this case the canonically conjugate momenta can be written as \cite{r27, r28}
  \begin{eqnarray}
  p_u&=&\frac{\partial L}{\partial \dot{u}}=\epsilon \dot{u} e^{2w+w_1-v}~~(\mbox{choosing}~~ w_{BD}=\frac{3}{2})\label{6.1}\\
 p_v&=&\frac{\partial L}{\partial \dot{v}}=-\frac{3}{2}e^{w_1}\dot{v}\label{6.2}
 \end{eqnarray}
 and consequently the Hamiltonian of the system takes the form
 \begin{equation}
 H=\frac{1}{2\epsilon}e^{v-2w-w_1}p^2_u-\frac{1}{3}e^{-w_1}p^2_v-v_0e^{(1+\frac{k}{2})v-w_1}e^{(4-k)w}\label{6.3}
 \end{equation}
 \begin{figure}[h]
 	\begin{minipage}{0.47\textwidth}
 		\centering \includegraphics[height=5cm,width=8cm]{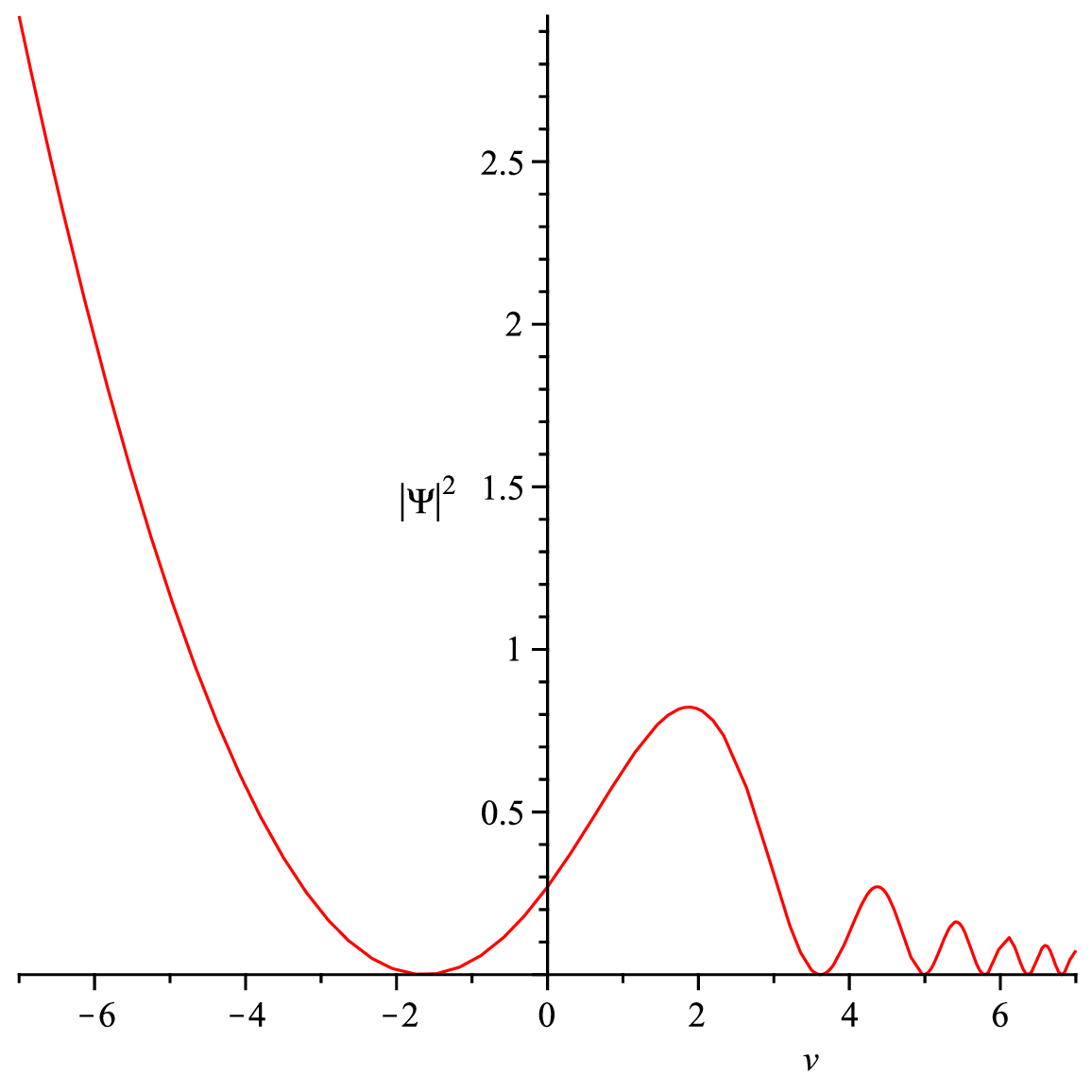}
 	\end{minipage}\hfill
 	\begin{minipage}{0.47\textwidth}
 		\centering \includegraphics[height=5cm,width=8cm]{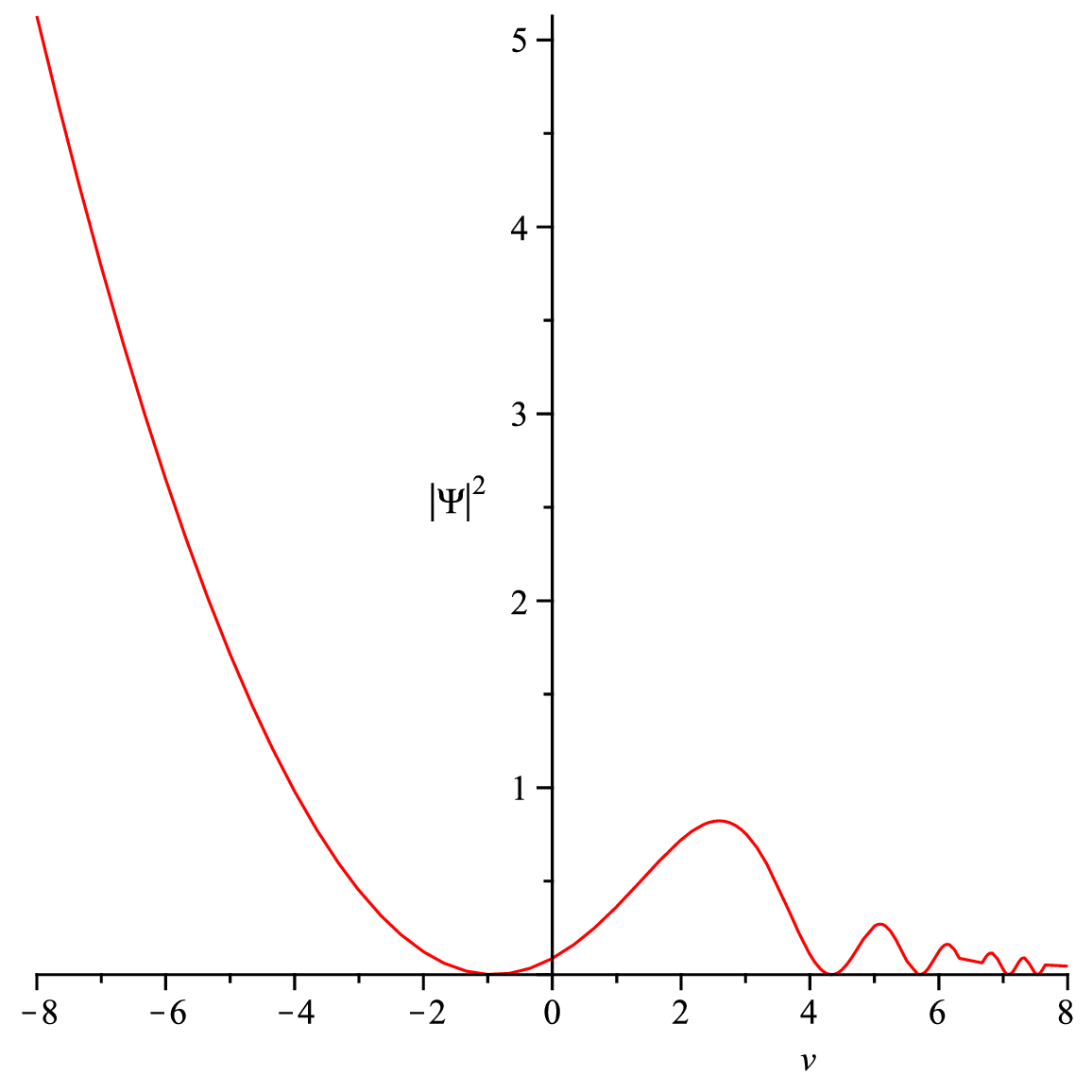}
 	\end{minipage}
 	\caption{shows the graphical representation of wave function $|\psi|^2$ (top left) when $k=0$, and (top right) when $k=-2$ against $v$ for Case I.}\label{f3}
 \end{figure}
 \begin{figure}[h]
 	\begin{minipage}{0.47\textwidth}
 		\centering \includegraphics[height=5cm,width=8cm]{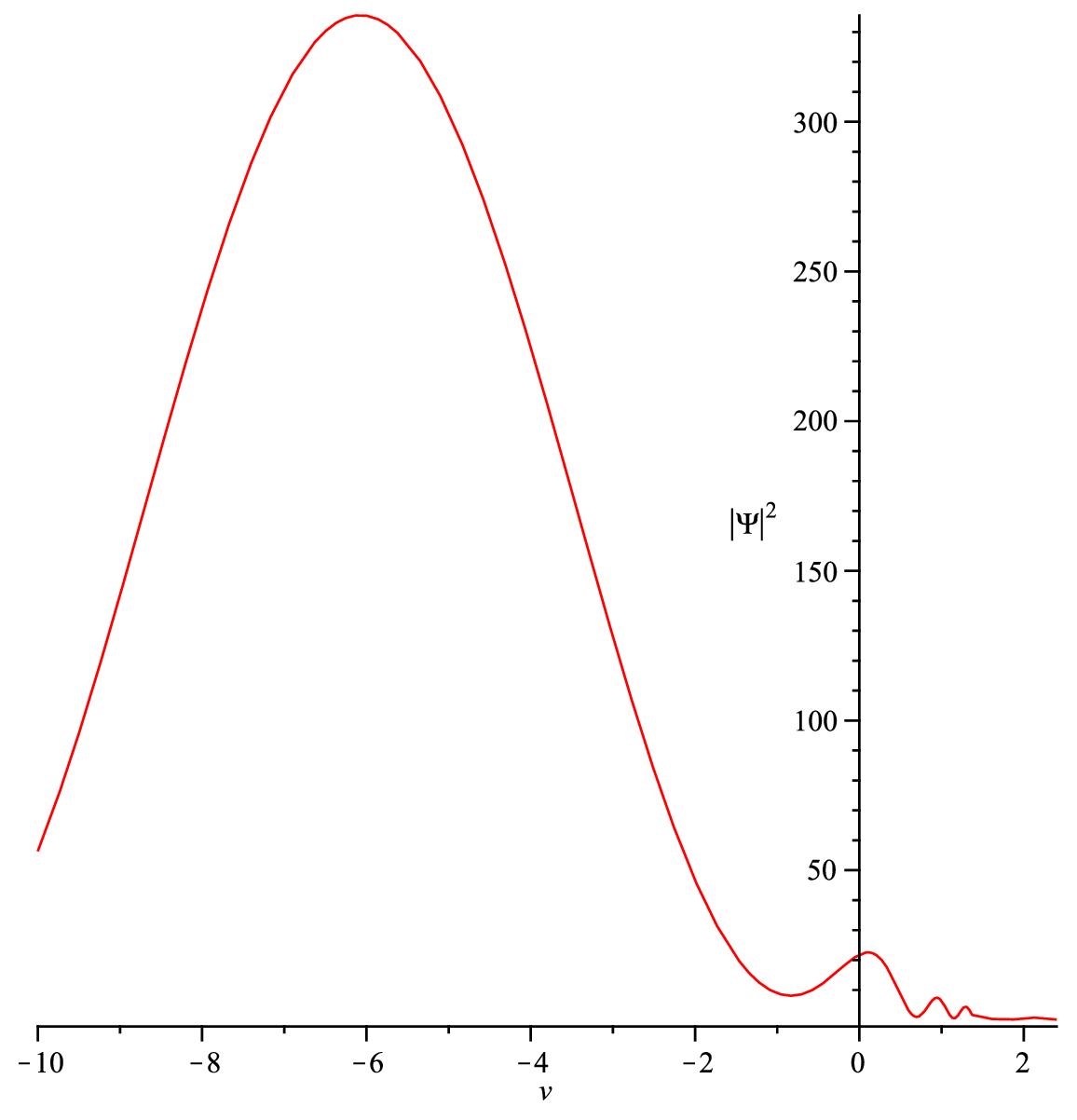}
 	\end{minipage}\hfill
 	\caption{Represents the wave function $|\psi|^2$ for case IV}\label{f4}
 \end{figure}
Hence the corresponding Wheeler-DeWitt equation can be written as
\begin{equation}
\bigg[-\frac{1}{2\epsilon}e^{v-2w-w_1}\frac{\partial^2}{\partial u^2}+\frac{1}{3}e^{-w_1}p^2_v\frac{\partial^2}{\partial v^2}-v_0e^{(1+\frac{k}{2})v-w_1}e^{(4-k)w}\bigg]\Psi(u,v)=0\label{6.4}
\end{equation} 
 where $\Psi$ is the wave function of the Universe for this model. Now, using the method of separation of variables $\Psi(u, v)=\Psi_1(u) \Psi_2(v)$ one can solve the above WD equation. Since $u$ is a cyclic coordinate, so the corresponding conjugate momenta ($p_u$) will be conserved in nature i.e.,
 \begin{equation}
 \epsilon \dot{u} e^{2w+w_1-v}=\mbox{conserved}=\Sigma_0 ~(\mbox{say})\label{6.5}
 \end{equation}
 The solution of the operator version of the equation (\ref{6.5}) gives the oscillatory part of the wave function as
 \begin{equation}
 \Psi_1(u)=e^{i\Sigma_0 u}\label{6.6}
 \end{equation}
using (\ref{6.6}) in the WD equation (\ref{6.4}), one gets the complete expression of the wave function as
\begin{equation}
\Psi(u, v)=e^{i\Sigma_0 u}\bigg\{c_1~ J \bigg(2\sqrt{3v_0} e^{3w}, \frac{\sqrt{6} \Sigma_0 e^{-w+\frac{v}{2}}}{\sqrt{\epsilon}}\bigg)+c_2~ Y \bigg(2\sqrt{3v_0} e^{3w}, \frac{\sqrt{6} \Sigma_0 e^{-w+\frac{v}{2}}}{\sqrt{\epsilon}}\bigg)\bigg\}
\end{equation} 
where $c_2, c_2$ are the integration constants and $J,~ Y$ are the usual Bessel and Neumann functions. Note that the quantum cosmological description for case IV is very similar to the above description for case I. Hence we have only given the graphical description. The graphical representation of $|\Psi|^2$ against $v$ has been shown in Fig.(\ref{f3}) and Fig.(\ref{f4}) for case I and case IV respectively. It is to be noted that in the subcases for each case the nature of the graph for $|\Psi|^2$ almost same and hence only two graphs are presented for the case I (with $k=0, k=-2$ respectively). For case I, $|\Psi|^2$ shows a damping oscillation at large $u$ while $|\Psi|^2$ increases as $v$ decreases to $-\infty$. Hence at zero volume (i.e., $v \rightarrow -\infty$) the probability measure has finite nonzero value and hence classical singularity can not be eliminated by quantum description. Similar is the situation in case IV. The oscillatory part of the probability measure gradually dies out as $v$ increases.
\section{Brief Summary and Conclusion} 
From cosmological point of view the present work deals with a modified gravity theory where the Brans-Dicke theory has been extended for two scalar fields which are minimally coupled among themselves. This model in Jordan frame can be viewed as the interacting two scalar field model with interaction in the potential term. However, Einstein frame, one may interpret this two scalar field model as quintom model and the second scalar field is of quintessence or phantom in nature depending on the choice of the parameter $\epsilon$ to be $1$ or $-1$. To judge the merit of the present cosmological model in FLRW space time, exact solutions to the modified Einstein field equations are essential. The challenge to solve these higher non-linear coupled set of differential equations Noether symmetry analysis has been used in this work. Though four distinct symmetry vectors are obtained but exact solution is possible only in two cases (for I and IV). In the other two cases cyclic coordinate is identified and Lagrangian is simplified to some extend but that is not sufficient to have the exact solutions. By analyzing the possible cosmological solutions it is found that the present model either describes the evolution starting from the earlier accelerated era to the present dark energy dominated expansion through the matter dominated era or the model describes the evolution from the decelerated matter dominated phase to the present late time accelerated epoch. to address the issue of initial big-bang singularity quantum cosmology has been formulated in Hamiltonian approach by constructing the WD equation. The wave function i.e., The solution of the WD equation has been used to have the probability measure and it has been plotted in Fig.(\ref{f3}) and Fig.(\ref{f4}). From the graphs it is evident that probability measure does not vanish at zero volume. Hence one may conclude that the classical singularity can not be eliminated by quantum description. 
\section*{Acknowledgments}
The author S. H thanks SVMCM (WBP231676269967) scholarship of higher education department Govt. of West Bengal, India and S.C. thanks FIST program of DST(SR/FST/MS-II/2021/101(C)).

\end{document}